# On-the-fly Autonomous Control of Neutron Diffraction via Physics-Informed Bayesian Active Learning

Austin McDannald[1,*], Matthias Frontzek[2], Andrei T. Savici[2], Mathieu Doucet[2], Efrain E. Rodriguez[3,4], Kate Meuse[5], Jessica Opsahl-Ong[6], Daniel Samarov[7], Ichiro Takeuchi[4,8], William Ratcliff[8,9,*], A. Gilad Kusne[1,8]

**Affiliations**
[1] Materials Measurement Laboratory, NIST, Gaithersburg, MD, 20899, USA
[2] Neutron Sciences Directorate, ORNL, Oak Ridge, TN, 37831, USA
[3] Department of Chemistry and Biochemistry, University of Maryland, College Park, MD, 20742 USA
[4] Maryland Quantum Materials Center, College Park, MD, 20742, USA
[5] Department of Computer Science, Cornell University, Ithaca, NY, USA
[6] Department of Computer Science, Rice University, Houston, TN, USA
[7] Information Technology Laboratory, NIST, Gaithersburg, MD, 20899, USA,
[8] Department of Materials Science and Engineering, University of Maryland, College Park, MD, 20742 USA
[9] NIST Center for Neutron Research, NIST, Gaithersburg, MD, 20899, USA
[*]Contact Author. Email: austin.mcdannald@nist.gov (A.M), william.ratcliff@nist.gov (W.R.)

**Abstract**

We demonstrate the first live, autonomous control over neutron diffraction experiments by developing and deploying ANDiE: the <u>A</u>utonomous <u>N</u>eutron <u>D</u>iffraction <u>E</u>xplorer. Neutron scattering is a unique and versatile characterization technique for probing the magnetic structure and behavior of materials. However, instruments at neutron scattering facilities in the world is limited, and instruments at such facilities are perennially oversubscribed. We demonstrate a significant reduction in experimental time required for neutron diffraction experiments by implementation of autonomous navigation of measurement parameter space through machine learning. Prior scientific knowledge and Bayesian active learning are used to dynamically steer the sequence of measurements. We show that ANDiE can experimentally determine the magnetic ordering transition of both MnO and $Fe_{1.09}Te$ all while providing a 5-fold enhancement in measurement efficiency. Furthermore, in a hypothesis testing post-processing step ANDiE can determine transition behavior from a set of possible physical models. ANDiE's active learning approach is broadly applicable to a variety of neutron-based experiments and can open the door for neutron scattering as a tool of accelerated materials discovery.

I. **Introduction**

Bayesian autonomous physical science – a novel, rapidly developing field, has the potential to revolutionize the scientific method at large by greatly accelerating the full knowledge capture process. Autonomous physical science places machine learning in control of the experimental system, at each iteration selecting and executing the most information-rich experiments. Some of the first embodiments of autonomous experiments in laboratory settings have taken the form of a mobile robotic chemist,[1] a self-driving laboratory for chemical synthesis,[2] an autonomous mechanical design system,[3] and an optimization tool for carbon nanotube growth.[4] With the recent demonstration of autonomous synchrotron diffraction,[5,6] the stage for AI-directed experiments has now expanded to national-laboratory beamline settings where allotted experimental beam time is a precious commodity.

The need for reduction in measurement time and resources is particularly acute in neutron scattering. Neutron diffraction is one of the few measurement techniques capable of directly observing magnetic order parameters. The magnetic order parameter describes the transition behavior from magnetically ordered to disordered states. Understanding how and at what temperature these magnetic ordering transitions occur is critical to discovering new magnetic materials. As a result, beamtime at neutron science facilities is highly prized. Yet this powerful technique is only available in a handful of facilities worldwide, with demand far outstripping supply.

Beamline experiments studying the order parameter typically perform exhaustive measurements defined *ad hoc*, with users defining a large sampling range and high sampling frequency in the hopes of not missing pertinent information. Currently, there is no universally agreed upon method for scheduling these measurements. This *ad hoc* scheduling is not too problematic when the diffraction signal is strong and the sample is well known, but is ineffective and potentially wasteful with samples with small signals (*i.e.* small magnetic moments, small crystals, or thin-films) or when little is known about the samples. One of the goals of this work is to algorithmically formalize the selection of measurements based on the statistical inference that can be drawn from them. In the traditional *ad hoc* schedule, once the full data is collected, an expert analyzes the data using computational tools to extract target parameters. In scattering experiments with high flux (e.g. synchrotron X-ray diffraction measurements) the high volume and velocity of data collected requires machine learning to rapidly elucidate pertinent scientific information.[7,8] However, the flux in neutron diffraction experiment is typically much lower and therefore collecting large data sets is both cumbersome and time consuming. This challenge is exemplified by an order parameter investigation, where the typical exhaustive method obtains far more measurements than required. For such challenges machine learning can serve to accelerate knowledge capture by optimizing each subsequent experiment on-the-fly through autonomous control. Here, we have developed the <u>A</u>utonomous <u>N</u>eutron <u>D</u>iffraction <u>E</u>xplorer (ANDiE) for self-directed probing of magnetic transition behavior with substantial improvement in measurement efficiency.

In this work, measurement selection is guided, on-the-fly, by a combination of active learning (AL) – a branch of AI dedicated to optimal experiment design (i.e. adaptive design)[9] – and probabilistic modeling/programming. A recent work has shown, with *in silico* autonomous campaigns, that Bayesian techniques could be used to determine the sequence of neutron scattering measurements.[10] In that work they show the efficacy of using Gaussian Processes to interpolate between pre-collected data. In the present work, we demonstrate – for the first time – live autonomous neutron scattering experiments using a physics-based approach that incorporates knowledge of neutron scattering and magnetic physics. The inclusion of physics-based knowledge into a Markov Chain Monte Carlo (MCMC) framework allows ANDiE to interpolate and extrapolate on-the-fly. This method allows the uncertainty to be captured and propagated throughout the process, from measurement data collection through model prediction. Furthermore, the prior physics encoded throughout ANDiE's MCMC framework restricts the analysis results to those that are physically realizable. This follows the design principles of the nascent field of Scientific AI (SciAI).[11] Such SciAI-driven autonomous experiments have the potential to significantly reduce the time needed for neutron diffraction experiments, alleviating demand on the instrument and the expert's time, thus allowing for more experiments at a given facility or the ability to perform otherwise impractical experiments. We show that ANDiE has successfully autonomously driven live neutron diffraction experiments to discover the Néel temperature ($T_N$) and to perform subsequent hypothesis testing for the temperature dependence of magnetic structure of the sample, all with about one fifth the number of measurements normally needed.

Active learning has roots as far back as the 18th century, when Laplace used AL to guide his study of celestial bodies.[12] Bayesian AL methods such as Bayesian optimization incorporate probabilistic modeling to output both predictions and associated uncertainties. These methods are particularly useful in guiding scientists in the lab or *in silico* to optimize unknown functions.[13–19] There is a pressing need for autonomous experimentation in the field of materials science in order to accelerate materials discovery.[20] As the driving force of autonomous systems, AL has been shown to be effective for optimizing materials processing conditions,[4,21,22] sample characterization,[6] and the composition of polymers and organic molecules for technological applications,[1,2,23] all using traditional AI optimization schemes. Many of these efforts in autonomous systems have been primarily focused on chemistry.[24–26] Reviews of the nascent field of autonomous systems for solid-state materials indicate significant promise for accelerating materials discovery and elucidating complex materials-property relationships.[27,28] By encoding prior physical knowledge into these AI tools, SciAI holds a significant advantage - further accelerating knowledge capture while also maintaining interpretability. For example, the closed-loop autonomous system for materials exploration and optimization (CAMEO) algorithm, which encodes knowledge of phase mapping and X-ray diffraction, was demonstrated to outperform non-SciAI method in accelerating materials optimization, and resulted in the discovery of a best-in-class phase-change memory material.[5]

ANDiE employs Bayesian SciAI to accelerate the determination of the magnetic order parameter from neutron diffraction measurements. The magnetic order parameter describes the behavior of the magnetic ordering transition. For example, an antiferromagnet with a first-order transition exhibits a discontinuity in order parameter at $T_N$, whereas one with a second-order transition exhibits an order parameter continuous in temperature with a discontinuity in the first derivative at $T_N$. When possible, preliminary magnetometry, heat capacity, or transport measurements can suggest that a magnetic ordering transition is suspected. In contrast, magnetic ordering is directly observable in neutron diffraction, as neutrons scattering off magnetic moments in the ordered state cause additional Bragg diffraction intensity. This allows neutron diffraction to study the magnetic ordering in materials that are otherwise difficult to measure with the aforementioned bulk characterization techniques (e.g., thin-film samples, or samples with impurity phases). Therefore, neutron diffraction is the most decisive measurement technique to determine the magnetic order parameter and $T_N$. Any preliminary characterization measurements could be used by ANDiE to inform the prior distributions of $T_N$, thereby allowing ANDiE to converge more quickly on informative data and the correct value of $T_N$. Furthermore, there are several well-known physics principles involved in this neutron diffraction process that are encoded in the presented algorithm for autonomously discovering the order parameter. These include the facts that diffraction intensities have Poissonian-like uncertainties, diffraction peaks are well described by Pseudo-Voigt profiles, and the magnetic component of the diffraction intensity is related to the square of the magnetization.[29] Importantly, the temperature dependence of the magnetization can follow a few models. A description of how the prior physical knowledge is encoded can be found in the Methods section.

We start the diffraction experiment by cooling from above the transition temperature to a base temperature of 5 K. At each subsequent iteration, physics-based uncertainty quantification guides the selection of subsequent isothermal diffraction measurements to maximize knowledge of magnetic structure peak parameters and their temperature dependence. The method combines an isothermal Bayesian inference step for the magnetic structure analysis with a second thermal Bayesian inference step for the temperature dependence analysis. While traditional active learning schemes would be able to select arbitrary temperatures in order to perform the optimal measurement, such a scheme is not possible for this task. First-order phase transitions are hysteretic; therefore, the measured $T_N$ depends on whether the sample was warming or cooling. Furthermore, due to this hysteretic nature of the order parameter below $T_N$, once the sample has been warmed above the base temperature it must be heated well above $T_N$ before it can be cooled to ensure that the order parameter is in the same state. This procedure would be prohibitively expensive, especially considering the goal of experiment is to discover the $T_N$ and therefore the heating step must be overestimated to confidently achieve the disordered state. This precludes the possibility of unconstrained exploration of temperature. We therefore restrict the acquisition function to only increase the temperature from base temperature. The measurement process repeats until the experiment temperature is well above inferred $T_N$ and no more information is gleaned by subsequent experiments. Figure 1 shows a diagram of the ANDiE scheme.

ANDiE was implemented for both the point-detector at the BT-4 beamline at the National Institute of Standards and Technology (NIST) Center of Neutron Research (NCNR) and for the Wide-Angle Neutron Diffractometer (WAND$^2$) at the HB-2C beamline at the Oak Ridge National Laboratory (ORNL) High Flux Isotope Reactor (HFIR). ANDiE was able to successfully control these instruments without human intervention and autonomously discover $T_N$ of both MnO and Fe$_{1.09}$Te powder samples while reducing the number of temperature steps to do so by a factor of ≈5 compared to traditional *ad hoc* scheduled experiments. Additionally, ANDiE performs hypothesis testing, identifying the correct physical model for the magnetic structure temperature dependence – in this case, either first-order, Ising-type second-order, or Weiss-type second-order. The models are ranked by likelihood and the top model is chosen. In this way, we have shown that ANDiE is capable of dramatically improving the efficiency of such neutron diffraction experiments.

## II. Materials and Methods
### A. Algorithm:

For the isothermal inference, we used model for the diffraction intensity in the *2θ* diffraction space range of interest constructed from two Pseudo-Voigt peaks and a constant background. The full model for the diffraction intensity in *2θ*-space is given by:

$$I = \zeta_{Mag} I(x - \omega_{Mag}, \alpha_{Mag}, \gamma_{Mag}) + \zeta_{Struct} I(x - \omega_{Struct}, \alpha_{Struct}, \gamma_{Struct}) + I_{Back} \quad (1.0)$$

where $\zeta_{Mag}$ and $\zeta_{Struct}$ are scaling factors, $\omega_{Mag}$ and $\omega_{Struct}$ are the peak locations in *2θ*-space, and $\alpha_{Mag}$, $\gamma_{Mag}$, $\alpha_{Struct}$, $\gamma_{Struct}$ are the Pseudo-Voigt peak shape parameters for the magnetic and structural peaks respectively, and $I_{Back}$ is the background intensity.

As discussed in the next section, we use the Weiss equation to predict the temperature dependence of the local magnetic moments during the autonomous experiment.[30] With the assumption that the diffraction peak shape does not strongly change over the course of the experiment, the maximum intensity of the magnetic diffraction peak is proportional to the integrated intensity. The full Weiss model for the temperature dependence of the magnetic diffraction intensity is then given by:

$$I(T) = \left(M_0 \underset{m}{\text{root}}\left[B_J\left(\frac{m}{T/T_N}\frac{3J}{J+1}\right) - m\right]\right)^2 + Bk \quad (2.0)$$

$$B_J(x) = \frac{2J+1}{2J}\coth\left(\frac{2J+1}{2J}x\right) - \frac{1}{2J}\coth\left(\frac{1}{2J}x\right) \quad (2.1)$$

where $\text{root}_m[]$ is the root finding operation of the expression in the square brackets with respect to $m$, $m$ is the reduced magnetization, $T$ is the temperature in Kelvin, $I(T)$ is the diffraction intensity, $T_N$ is the magnetic transition temperature, $J$ is the quantum total angular momentum, $M_0$ is a scaling parameter proportional to the maximum spontaneous magnetization, $B_J(x)$ is the Brillouin function, and $Bk$ is the background intensity. Note that the $M_0$ in eq. 2.0 is a composite of the maximum spontaneous magnetization and the square root of the unknown proportionality constant between the diffraction intensity and the square of the magnetic moments.

To select the next temperature, temperature values are explored (with a step size of 0.5 K) to identify the next temperature where the confidence interval of the model exceeds a threshold relative the Poissonian-like uncertainty predicted by the mean of the model. Limiting the active learning scheme to increasing temperature avoids any hysteretic effects. Once the temperature is above the upper confidence bound of $T_N$, the confidence interval of the model no longer depends on temperature (as the background is the only parameter left to fit). Large temperature steps above $T_N$ are then taken.

Once the full data set is collected in the autonomous experiment, ANDiE performs a post-processing hypothesis testing to determine which of the models mentioned in the Discussion section is the most likely. The first-order model is given by:

$$I(T) = \left(\frac{K}{2}\left(1 - erf\left(\frac{T-T_N}{\sigma\sqrt{2}}\right)\right)\right)^2 + Bk \qquad (3.0)$$

where $K$ is the intensity scaling constant, and $\sigma$ is the full-width at half maximum of the Gaussian convolution of the step-function which is used to describe the width of the transition. Lastly the Ising model is given by:

$$I(T) = \left(M_0\left(1 - \frac{T}{T_N}\right)\right)^{4\beta} + Bk \qquad (4.0)$$

where $\beta$ is the critical exponent. Note that equation 4.0 is only valid near $T_N$, which we have used $T > 0.5 T_N$ to enforce.

ANDiE performs the inference using each of the three models. Note that the inference with the Ising model infers $T_N$ and therefore also the range in which the Ising model is valid. To ensure a fair comparison of the log-likelihoods between each of the three models, once the Ising model inference is complete, we re-perform the inference of the first-order and Weiss models using only the data points that fall within the valid range of the Ising model. If the Ising model is not the most likely given the data points within that range, we compare the first-order and Weiss models using the inference on the entire temperature range acquired by the autonomous experiment.

All parameters of each model are initialized with prior truncated normal distributions based on the physical limitations (*e.g.*, $T_N$ cannot be negative) and estimates from experts. ANDiE uses the DREAM sampler[31] to perform the MCMC inference. For each of the models (in both the isothermal and thermal inference), we use a Gaussian likelihood around the prospective curve to determine the probability of observing the data given the models. The widths of these likelihood distributions are determined by model and the instrument uncertainties,[32,33] which captures the highly heteroscedastic nature of these Poissonian-like processes. The autonomous analysis of the neutron diffraction data shown here was enabled by data pipeline that automatically reduces neutron event data into spectra[34] using the Mantid framework.[35]

The full algorithm was written in Python and implemented in a Jupyter notebook that analyzes the diffraction patterns, selects the next temperature, and communicates with data acquisition, without human intervention. The ANDiE notebooks used during the autonomous experiments are available at: https://github.com/usnistgov/ANDiE-v1_0. The BUMPS library[36] was used for the MCMC functions with the DREAM sampler[31]. For the thermal inference a numerical root-seeking algorithm from the sci-kit learn library[37] is used to solve for the root of the Weiss equation at each step in the MCMC chain.

B. **Experimental Set-up:**

Diffraction experiments were performed at the WAND$^2$ HB-2C beamline at HFIR at ORNL using a wavelength of 1.4828 Å. Initial algorithm development was using experiments performed at the BT-4 beamline at the NCNR at NIST. The MnO powder was purchased from Sigma Aldrich (Cat. # 377201)*. A description of the synthesis details of the Fe$_{1.09}$Te powder sample can be found in.[38] Both MnO and Fe$_{1.09}$Te powder samples were measured in Vanadium cans sealed under He-atmosphere. To reach low temperatures a top-loading closed cycle refrigerator with a variable temperature insert (VTI) with He-exchange gas was used.

III. **Results and Discussion**
A. **Autonomous Discovery of Magnetic Transition Behavior of MnO:**

In this first demonstration of an autonomous research neutron diffraction system, we initially consider the well-studied material MnO to ensure ANDiE can reproduce known results. In the next section we consider the more challenging material Fe$_{1.09}$Te, which, as ANDiE discovers, has a sharp first-order transition. These studies demonstrate the robustness of ANDiE and future studies can therefore confidently use ANDiE to study materials where the magnetic transition behavior is unknown. There are some materials where the magnetic propagation vectors - and therefore diffraction peak positions in $2\theta$-space - are strong functions of temperature.[39,40] Additionally, the intensity of the magnetic contribution to the diffraction pattern is also a strong function of temperature, especially across the ordering temperature (compare Figure 2a to Figure 2b). As a result, the isothermal model parameters can change dramatically as the temperature-dependence experiment progresses. This motivates the need for a reliable algorithmic platform capable of capturing such diverse behavior. Bayesian inference provides a robust, probabilistic method to describe the material at any one temperature and across temperatures. Bayesian inference allows one to utilize prior knowledge to improve data analysis and prediction and it provides a framework for uncertainty quantification and propagation. In contrast to simpler methods such as least squares fitting, this Bayesian framework allows the parameters to be inferred from the data with more accurate uncertainties. Data can be input with uncertainty bounds, and target parameters are output as probability distributions with expected value and uncertainty. In particular, ANDiE uses MCMC-based Bayesian inference to extract the magnetic component from each isothermal diffraction measurement. The use of MCMC inference for global optimization ensures high confidence in peak parameter determination despite the large range of potential parameters values. MCMC is particularly well suited to avoiding the myriad of local minima present in diffraction data.[41] In contrast, other optimization schemes, such as the Levenberg-Marquardt algorithm, can perform well only when the initialization is close to the global minimum and can diverge otherwise. Additionally, MCMC allows us to encode prior physics knowledge such as the Poissonian-like counting statistics of the measured intensities, thereby accounting for the highly heteroscedastic nature of the intensity as a function of diffraction angle (and of temperature, as shown in the next section). Furthermore, MCMC prior estimates of the parameters can be included, *i.e.* nuclear peak positions from previous X-ray diffraction measurements, or information from previous reports in the literature. The active learning process begins with a previously identified range of interest for $2\theta$. For MnO we started with the detector $2\theta$ range of 28.0° to 37.0°, which includes the (111) nuclear peak and the nearby $\left(\frac{3}{2}\frac{1}{2}\frac{1}{2}\right)$ magnetic peak. ANDiE then infers probability distributions for the peak shape parameters including the locations, heights, half-widths at half maximum for both the Gaussian and Lorentzian components of Pseudo-Voigt peaks, as well as a background term. Figure 2 shows the results of this inference at 5.0 K and at 129.5 K.

ANDiE uses the inferred peak parameter distributions at each isothermal measurement to predict the temperature dependence of the diffraction's magnetic component. The magnetic component of neutron diffraction intensity is related to the square of the magnetic moment.[29] Several models can describe the temperature dependence. In this work we consider a first-order phase transition model and the Ising and Weiss second-order phase transition models. The first-order phase transition model is an error function - a step function convolved with a narrow Gaussian function. Because this model only has a non-zero slope near $T_N$, predictions made by inference are not informative for selecting subsequent temperature steps, *i.e.* there is no indication that the experiment temperature is approaching $T_N$ until it is within a few Kelvin (*i.e.* within the Gaussian convolution). Therefore, even if the material being studied is suspected to have first-order transition behavior, using a first-order model is not appropriate during the autonomous experiment. Instead, ANDiE uses a second-order model to make predictions during the autonomous experiment. A Bayes factor test is then used once all the data has been collected to determine if the material exhibits first-order behavior.

The second-order phase transition models do have non-zero slopes far below $T_N$ and therefore can be used to predict an appropriate temperature step. However, the Ising model is only valid near $T_N$, in the range $0.5T_N < T < T_N$,[42] whereas the Weiss model is valid across the entire temperature range below $T_N$. Therefore, regardless of the material being studied and the suspected behavior, ANDiE uses the Weiss model during the autonomous experiment to drive the data acquisition, and then in post-processing it determines the most appropriate model with the Bayes factor. Example curves of each of these models are shown in Figure 3.

For these reasons – regardless of the material being studied – ANDiE uses the Weiss model to select the temperature steps of subsequent measurements, propagating knowledge from low to high temperatures. This physics-informed approach has several advantages over a more generic power law fitting or a surrogate ML model such as a Gaussian process. Firstly, the Weiss model constrains ANDiE to only physically meaningful solutions, *i.e.* positive temperature, positive intensity, and monotonic temperature dependance. Secondly, we demonstrate that with the Weiss model, ANDiE focuses measurements in the most informative regions: at low temperature and surrounding the transition temperature. Furthermore, ANDiE uses a minimum number of measurements to properly characterize the curvature and background outside these regions. We found this to be true regardless of the actual materials behavior and using only a broad prior estimation of the $T_N$. For materials with second order transitions the Weiss model is flexible enough to drive the data collection to the informative temperatures. In the case of a truly abrupt discontinuous first-order transition, discovering such a step-function is a daunting task that can only be solved iteratively with several cooling and warming cycles and ANDiE could be implemented to autonomize these iterations. However, if there is a perturbation from ideal first-order behavior (*i.e.* from short-range order or the like) as is the case for many materials, the flexibility of the Weiss model allows ANDiE to collect more data near the $T_N$. In this way, ANDiE uses the Weiss model to discover the $T_N$ (with enough data to determine the transition behavior) from a single warming cycle.

Because neutron diffraction intensity obeys Poissonian-like counting statistics the process is highly heteroscedastic, meaning that the uncertainty in the signal is highly non-uniform across the search domains. The uncertainty (as estimated by the standard deviation of a Gaussian distribution using the continuous approximation) of the diffraction intensity is related to the square root of the diffraction intensity. Common acquisition functions do not account for heteroscedasticity and tend to over-emphasize regions of high intensity, unnecessarily acquiring more data in these regions. ANDiE therefore compares the confidence interval of the model (a value dependent on the number of measurements) to the uncertainty predicted from the intensity extrapolation. ANDiE increases the temperature until that ratio is above some threshold which we call the Bravery factor. The model variance is a measure of how well known the intensity is at each temperature given the data that has been measured. The predicted uncertainty is a measure of how much we should expect to know about the value of intensity if a measurement is performed at that temperature. The ratio of these two values represents how informative that measurement will be to the model. Setting the Bravery factor determines a threshold on this ratio, above which measurements are considered useful. Temperatures where the ratio is below the Bravery factor can be safely skipped, and the experiment temperature can be increased until that threshold is reached. The Bravery factor therefore represents the user's risk tolerance and can change depending on the purpose of the experiment. If little is known about the material, a high Bravery Factor might be appropriate to explore the space quickly. If, however, the goal of the experiment is to fine-tune the measurement of $T_N$ then a smaller Bravery Factor might be appropriate so as to only take small temperature steps (*i.e.*, measurements that are only moderately informative are still useful).

Figure 4 shows how ANDiE performed for the autonomous discovery of $T_N$ of MnO. ANDiE chooses small temperature steps in the beginning of the autonomous experiment as there is little data to infer the temperature dependence. As more data is acquired, ANDiE takes larger temperature steps until it approaches $T_N$. Near $T_N$, the steep slope of the model naturally causes wide confidence intervals of the inference, and more data is acquired in the region. In this way ANDiE skips uninformative temperatures and quickly converges on $T_N$. After 14 temperature steps ANDiE inferred that the experiment temperature was above $T_N$. In this region the selection of further data points is arbitrary, and several measurements were taken at 10 K steps. The results of this inference at several stages are shown in Figure 4. After 16 temperature steps ANDiE reached the stopping criteria for the experiment. ANDiE quickly converged on the most likely parameters. As mentioned before there is no universally agreed method for the traditional *ad hoc* scheduling, which is determined by the intuition of the researcher and is particularly difficult with small signals (i.e. from materials with small magnetic moments, small crystal samples or thin-film). For the sake of comparison, an informed *ad hoc* schedule might take 0.5 K steps within 10 K of prior guess of $T_N$, 2 K steps within 20 K of the prior guess of $T_N$, and 5 K steps otherwise, for a total of 74 temperature steps. ANDiE therefore reduces the number of temperature steps required for the experiment by a factor of ≈5.

After the autonomous experiment reaches the stopping criteria, ANDiE performs hypothesis testing to determine which of the models considered herein are more likely. Since the Ising model is only valid near $T_N$, it determines the range over which the model likelihoods will be compared. Thus, ANDiE performs inference with the Ising model first. Inference is then performed for the other models over the same temperature range. If the Ising model is not the most likely over the appropriate range, then ANDiE compares the likelihoods of the first-order model and Weiss model over the entire data set. Figure 5 shows the result of the Ising model inference, determined to be the most likely model, with an estimated $T_N$ of 120.81(56) K. Table 1 summarizes the results for all models. The uncertainties in the prediction of $T_N$ reflect the confidence of the model in that parameter given the data points. These confidence intervals represent an uncertainty in the parameter only insofar as the model is physically applicable. For example, in the case of MnO, the Weiss model is not likely physically meaningful. This is therefore also true of uncertainty in the $T_N$ as derived from the Weiss model for that sample. The fact that the first-order and Weiss models have such unphysically low uncertainty in this prediction shows that adjusting $T_N$ further will not improve the fit to the data. This is also reflected in the large negative log-likelihoods of these models showing that they are not appropriate for the data. In contrast, the Ising model is appropriate, as evidenced by the higher log-likelihood. Therefore, the confidence in the $T_N$ parameter from the Ising model inference is a good measure of the uncertainty in $T_N$. Following this, ANDiE concludes that MnO is an Ising-type antiferromagnet with a $T_N$ of 120.81(56) K, consistent with the literature.[42–44]

B. **Autonomous Discovery of Magnetic Transition Behavior of Fe$_{1.09}$Te:**

Having validated its effectiveness on determining the magnetic transition of MnO, a well-studied material with a second-order transition, ANDiE was then implemented on the more challenging Fe-Te system. Fe$_{1+x}$Te has complicated magnetic behavior as a function of the interstitial iron, i.e. the $x$ in the chemical formula.[45] Below ≈11 % interstitial Fe, there is a first-order phase transition to an antiferromagnetic phase. $T_N$ of this transition in Fe$_{1+x}$Te ranges from 70 K at $x$ = 0 % to 52 K at $x$ = 11 %. Precise determination of $T_N$ for Fe$_{1.09}$Te is a challenging task since abrupt step-like first-order transition could occur over a wide range of temperatures. Indeed, Figure 6a shows sharp this transition is in the diffraction intensity of the magnetic $\left(\frac{1}{2}\ 0\ \frac{1}{2}\right)$ reflection of Fe$_{1.09}$Te at 69.436(55) K, as acquired by an *ad hoc* measurement schedule (carried out as a separate experiment after the autonomous run). ANDiE, using the Weiss-type transition model as a prior, discovered this transition in only 14 measurements. This is an improvement over the *ad hoc* schedule by a factor of ≈4. A discussion of the time savings, computational considerations, as well as a video capturing the evolution of inference as the autonomous experiment was performed are available in the supplemental material. We note that the code currently used to implement the ANDiE is developmental and while it is robust enough to demonstrate the autonomous decision making, future work could use parallel computing to speed computation (See discussion of computational time in Supplemental Section II). After the data is collected, ANDiE then performs the model comparison between the Ising, Weiss, and first-order models (as shown in Figures 6b – 6f). It can be seen that while the Ising model (Figure 6d) has reasonably good agreement with the data, the first-order model over the same temperature range (Figure 6f) provides a better description of the behavior. This is especially evident in the region near the $T_N$. This low intensity data has low uncertainty owing the Poisonian-like statistics. As a result of propagating this measurement uncertainty through this Bayesian framework, a few counts deviation between the model and the measurement at low intensity is far less likely than a few counts deviation at high intensity. Therefore, deviations between the model and the data are more heavily penalized at low intensity in the calculation of the model likelihood. The summary of the log-likelihoods and predicted transition temperatures used for the model selection is presented in Table 1. Here, it is worth noting that, as was the case above, the confidence intervals from the models are a good measure of the uncertainty only when the model is physically appropriate. Considering the data from full temperature range, the high log-likelihood of the first-order model indicates that this model is the most likely. ANDiE therefore correctly inferred first-order transition behavior with $T_N$ of 68.58(16) K, which agrees with that measured by the *ad hoc* schedule. This magnetic ordering behavior is similar to what is expected from earlier reports of the Fe$_{1+x}$Te phase diagram.[45] ANDiE is therefore able to discover the behavior of the magnetic order parameter in very few measurements even when the actual behavior is far from the prior estimation.

These results show that ANDiE is capable of autonomously discovering the $T_N$ of a material and performing basic model selection in the first live autonomously driven neutron diffraction experiments. This demonstration goes beyond a simple proof-of-concept by making a discovery of $T_N$ and transition behavior on a previously under-studied material. Furthermore, we have demonstrated the efficacy of a single pass of ANDiE, which can accomplish the goal of discovering $T_N$ to less than a degree and perform simple model selection. An accurate determination of critical exponents is beyond the scope of this prototyping work, but ANDiE could be generalized to accomplish that task. Future work can include conditions for cooling back to base temperature for subsequent runs of ANDiE with updated priors, in order to perform more difficult discovery tasks.

**Table 1.** Thermal inference results. The predicted Néel temperature ($T_N$) and model log-likelihoods are used in the post-processing hypothesis testing for the autonomous neutron diffraction study of MnO and $Fe_{1.09}Te$ with WAND$^2$ at HB-2C at HFIR at ORNL. Note that the Ising model is only valid near $T_N$. The uncertainty in the prediction of $T_N$ reflects the confidence of that model in that parameter given the data. These confidence intervals are good measures of the uncertainty only when the models are physically appropriate – as evidenced by the higher log-likelihoods. Note that uncertainty is presented in compact notation where (##) represents the uncertainty in the last two digits of the value.

| Material | Model | Full Range | | $T > 0.5 T_N$ | |
| --- | --- | --- | --- | --- | --- |
| | | $T_N$ (K) | Log Likelihood | $T_N$ (K) | Log Likelihood |
| MnO | First-Order | 114.49(12) | -2722 | 119.204(86) | -742 |
| | Weiss | 127.948(47) | -427 | 127.954(52) | -389 |
| | Ising | ----- | ---- | **120.81(56)** | **-52** |
| $Fe_{1.09}Te$ | First-Order | **68.58(16)** | **-64** | 69.10(18) | -21 |
| | Weiss | 70.000(16) | -348 | 70.000(17) | -287 |
| | Ising | ----- | ---- | 65.04(58) | -49 |

## IV. Conclusion

We have developed the autonomous neutron diffraction explorer (ANDiE), a system for controlling neutron diffraction experiments for the discovery of the magnetic transition temperature. The system presented here provides a Bayesian approach to selecting the experiment temperatures which not only provides probabilistic predictions but also encodes the relevant physics to the problem at hand. We have demonstrated the versatility of ANDiE which is capable of discovering the magnetic transition temperature of material systems with differing magnetic behaviors despite always driving the acquisition of data with the Weiss model. Even though this model might not be the expected behavior of the material being studied, it is useful in choosing the next temperatures to efficiently discover the transition temperature. ANDiE can accelerate the data acquisition by reducing the number of temperature steps by nearly a factor of 5 and can subsequently perform hypothesis testing to determine the governing physical principles of the transition. The hypothesis testing after the data is acquired correctly identified the Ising-type transition in MnO at 120.81(56) K. Even in the more challenging case of $Fe_{1.09}Te$ with an abrupt step-like first-order transition, ANDiE was able to efficiently drive the experiment; requiring only 14 measurements to discover the first-order transition behavior at 68.52(16) K. As currently implemented, ANDiE compares the likelihood between the three previously discussed models. Extending hypothesis testing to additional user-determined models is a straightforward task. Furthermore, the methods implemented in ANDiE can be easily expanded to a variety of neutron-based experiments. We expect the experiment speedup to increase with the dimensionality of the experiment such as implementing the active learning in the diffraction angle space on point-detector instruments such as the BT-4 at NCNR for rapid search of diffraction peaks. Similarly, using our active learning scheme in the applied magnetic field space would reduce the number of measurements for those experiments. The approach used by ANDiE is further generalizable to other measurements (such as X-ray diffraction or functional property measurements) where a generally applicable physical model can be used to efficiently navigate costly experimental conditions. The autonomous system presented here exemplifies the potential of rapid neutron scattering experiments for accelerating materials discovery.

**Acknowledgments**

We acknowledge assistance from Natasha Shmunis and Stephen Pheiffer from the NICE instrument control team.

We also acknowledge the help of Ray Gregory, Rich Crompton, Rob Knudson IV, and Jim Kohl for the setup of data acquisition control on WAND.
	A portion of this research used resources at the High Flux Isotope, a DOE Office of Science User Facility operated by the Oak Ridge National Laboratory.

**Funding:**
	U.S. Department of Commerce (DOC), National Institute of standards and Technology (AM, AGD, WR, DS)
	National Institute of Standards and Technology, National Science Foundation, Center for High Resolution Neutron Scattering Agreement No. DMR-2010792. (JO, KM)
	University of Maryland, NIST Cooperative Agreement 70NANB17H301. (IT)
	U.S. Department of Energy (DOE), Office of Science Grant DE-SC0016434 (EER)

**Author contributions:**
Conceptualization: IT, AGK, WR, AM
Methodology: AM, WR, AGK, MF, DS
Verification: JO, KM
Sample Preparation: EER, MF
Experiment Operation/Implementation: MF, ATS, MD, WR, AM, AGK
Supervision: IT, AGK, WR
Writing—original draft: AM
Writing—review & editing: AM, AGK, IT, WR, MF, ATS, MD, EER

**[*] Disclaimer: The identification of any commercial product or trade name does not imply endorsement or recommendation by the National Institute of Standards and Technology.**

**Data and materials availability:**
All of the code needed to implement ANDiE is available at:
https://github.com/usnistgov/ANDiE-v1_0.
All data are available at:
DOI: 10.18434/mds2-2449

**Figures and Tables**

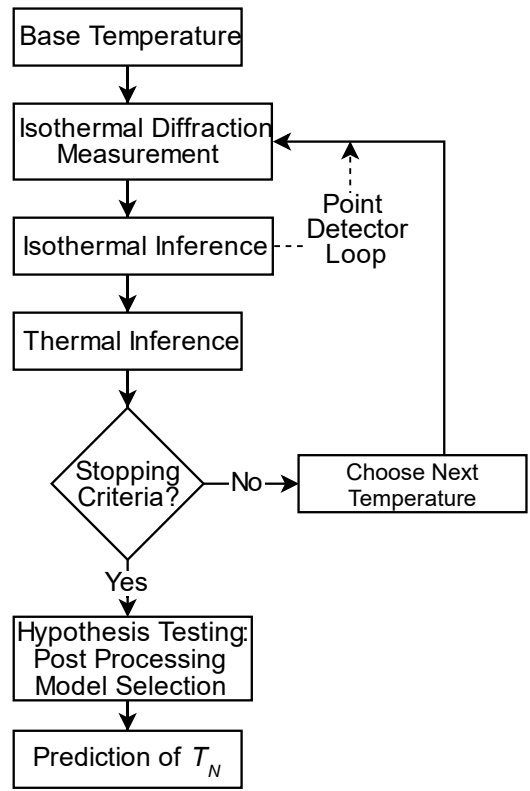

**Fig. 1:** A diagram of the algorithm used the <u>A</u>utonomous <u>N</u>eutron <u>D</u>iffraction <u>E</u>xplorer (ANDiE). ANDiE autonomously drives live neutron diffraction experiments to discover the Néel temperature ($T_N$) and to perform subsequent hypothesis testing for the temperature dependence of magnetic structure. The solid lines show how ANDiE was implemented on the Wide-Angle Neutron Diffractometer (WAND$^2$) at the HB-2C beamline at the Oak Ridge National Laboratory (ORNL) High Flux Isotope Reactor (HFIR), while the dashed line shows the additional active learning loop implemented on the BT-4 beamline at the National Institute of Standards and Technology (NIST) Center of Neutron Research (NCNR).

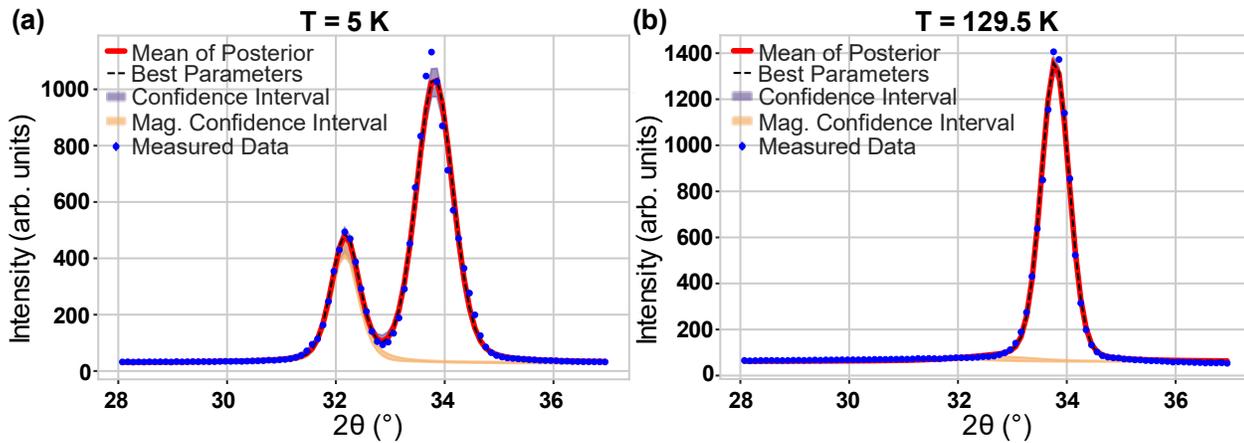

**Fig. 2:** Isothermal inference for MnO. Isothermal inference was performed on MnO diffraction data in the range of interest at the experiment temperature ($T$) of (a) 5.0 K and (b) 129.5 K. The magnetic confidence interval in orange shows the confidence interval of the magnetic component of the isothermal model. Note how the magnetic peak parameters $\left(\frac{3}{2}\frac{1}{2}\frac{1}{2}\right)$ near 32.18° changes between the temperatures. The global optimization MCMC algorithm infers an appropriate profile despite the large changes. Error bars on the measured data points in blue represent one standard deviation and are smaller than the symbol size.

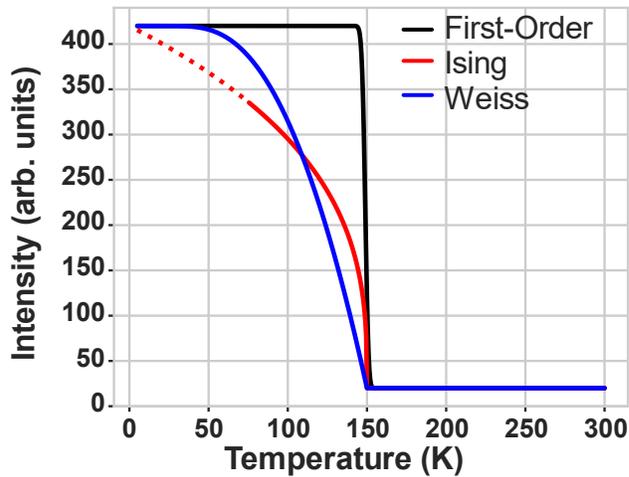

**Fig. 3:** Example models for thermal inference. There are several models for the temperature dependence of the magnetic component of the neutron diffraction intensity. The black curve shows the first-order model. The Ising model is shown in red, where the dashed region is outside the range of validity of this model. The Weiss model is shown in blue.

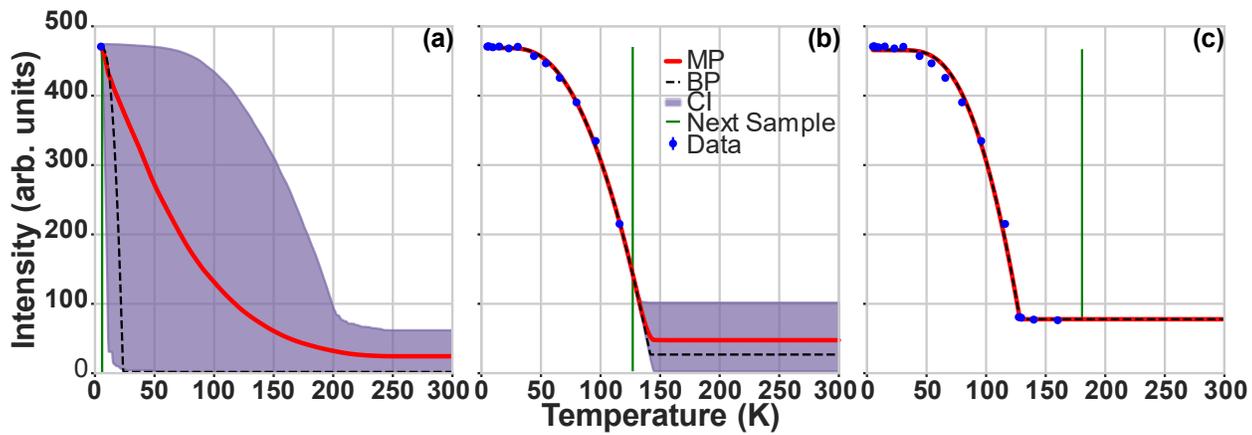

**Fig. 4:** Thermal inference snapshots for MnO. The thermal inference step was performed during the autonomous experiment for temperature dependence of the MnO magnetic $\left(\frac{3}{2} \frac{1}{2} \frac{1}{2}\right)$ reflection using the Weiss model. Results are shown as determined after (a) one measurement, (b) after 12 measurements near the Néel temperature ($T_N$), and (c) after 16 measurements at the end of the autonomous experiment. The mean of the posterior curves (MP) of the inference for each model is shown in red. The best parameters (BP) for each model are shown in the black dashed curves, while the confidence intervals (CI) are shown as the gray envelope. The vertical green line in each part shows the next temperature the algorithm selected to measure next. The error bars on the measured data points are smaller than the makers shown in blue, and in (c) the confidence interval of the model is smaller than the linewidth of the mean of the posteriors.

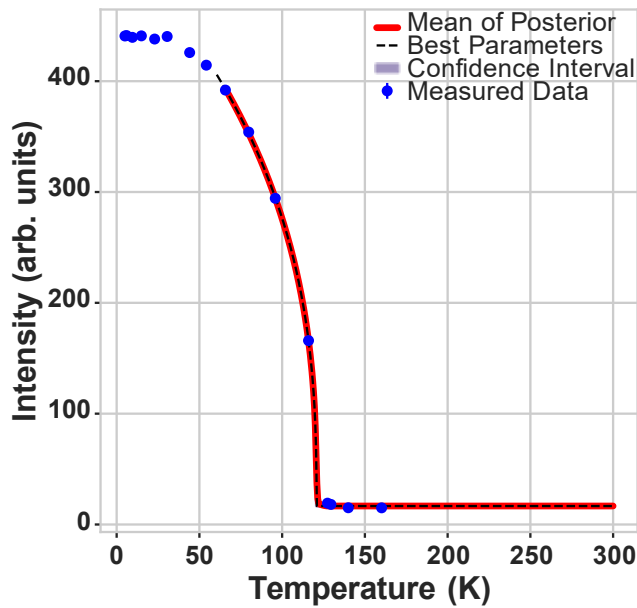

**Fig. 5:** Ising model hypothesis testing. Ising model inference was performed on the temperature dependence of the MnO magnetic $\left(\frac{3}{2} \frac{1}{2} \frac{1}{2}\right)$ reflection as performed the post-processing hypothesis testing step and determined to be the most likely model. The error bars on the measured data points are smaller than the makers, and the confidence interval of the model is smaller than the linewidth of the mean of the posteriors.

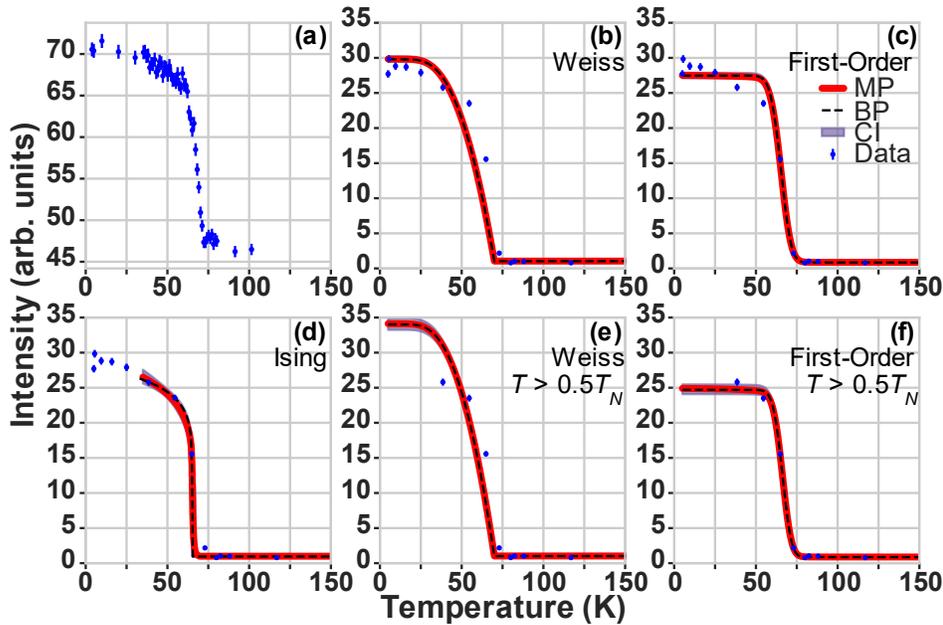

**Fig. 6:** Autonomous measurement and hypothesis testing vs. ad hoc schedule. The intensity of magnetic $\left(\frac{1}{2}\ 0\ \frac{1}{2}\right)$ reflection of $Fe_{1.09}Te$ as determined from the isothermal inference are shown in blue. The *ad hoc* schedule is shown in (a). The hypothesis testing was performed using the inference of the (b) Weiss, (c) first-order, (d) Ising models, respectively, on the autonomously acquired measurements. The (e) Weiss and (f) first-order models were re-trained on the data where experiment temperature greater than one half the estimate of the Néel temperature ($T_N$) from the Ising inference (i.e., $T > 0.5T_N$). The mean of the posterior curves (MP) of the inference for each model is shown in red. The best parameters (BP) for each model are shown in the black dashed curves, while the confidence intervals (CI) are shown as the grey envelope. The Weiss model was used to drive the autonomous experiment. The first-order model was determined to be the most likely for this transition of $Fe_{1.09}Te$. The error bars on the measured data points shown in blue are smaller than the makers, and the confidence interval of the model is smaller than the linewidth of the mean of the posteriors.